\documentclass{pasa}%

\usepackage{graphicx}
\usepackage{xcolor}
\usepackage{hyperref}
\usepackage{amsmath,amstext}
\usepackage[figure,figure*]{hypcap}
\usepackage{newtxmath} %use times font for math

\title[Breakthrough Listen Parkes Data Recorder]{The Breakthrough Listen Search for Intelligent Life: Wide-bandwidth Digital Instrumentation for the CSIRO Parkes 64-m Telescope}

\newcommand{\UCB}{Department of Astronomy,  University of California Berkeley, Berkeley CA 94720}
\newcommand{\SSL}{Space Sciences Laboratory, University of California, Berkeley, Berkeley CA 94720}
\newcommand{\SWIN}{Centre for Astrophysics \& Supercomputing, Swinburne University of Technology, Hawthorn, VIC 3122, Australia}

\newcommand{\OXF}{Astronomy Department, University of Oxford, Keble Rd, Oxford, OX13RH, United Kingdom}
\newcommand{\NIJ}{Department of Astrophysics/IMAPP,Radboud University, Nijmegen, Netherlands}
\newcommand{\ATNF}{Australia Telescope National Facility, CSIRO, PO Box 76, Epping, NSW 1710, Australia}
\newcommand{\NECT}{School of Science and Technology, Hellenic Open University, Parodos Aristotelous 18, Patra 26 335, Greece}

\newcommand{\dr}{BLPDR}

\newcommand{\refsec}[1]{Sec.~\ref{#1}}
\newcommand{\reffig}[1]{Fig.~\ref{#1}}
\newcommand{\reftab}[1]{Tab.~\ref{#1}}

\author[D. C. Price et al.]{
Danny C. Price$^{1,2}$\thanks{corresponding author: dancpr@berkeley.edu},
David H.\ E.\ MacMahon$^1$,
Matt Lebofsky$^1$,
Steve Croft$^1$,
David DeBoer$^1$,
J. Emilio Enriquez$^{1,3}$, 
Griffin S. Foster$^{1,4}$,
Vishal Gajjar$^{1,5}$,
Nectari Gizani$^{1,6}$,
Greg Hellbourg$^1$,
Howard Isaacson$^1$,
Andrew P. V. Siemion$^1$,
Dan Werthimer$^{1,5}$,
James A. Green$^7$,
Shaun Amy$^7$,
Lewis Ball$^7$,
Douglas C.-J. Bock$^7$,
Dan Craig$^7$,
Philip G. Edwards$^7$,
Andrew Jameson$^2$,
Stacy Mader$^7$,
Brett Preisig$^7$,
Mal Smith$^7$,
John Reynolds$^7$,
John Sarkissian$^7$

\affil{$^1$\UCB}%
\affil{$^2$\SWIN}
\affil{$^3$\NIJ}
\affil{$^4$\OXF}
\affil{$^5$\SSL}
\affil{$^6$\NECT}
\affil{$^7$\ATNF}

}%

\jid{PASA}
\doi{10.1017/pas.\the\year.xxx}
\jyear{\the\year}

\usepackage{aas_macros}
\usepackage{hyperref} 
\hypersetup{colorlinks,citecolor=blue,linkcolor=blue,urlcolor=blue}

\hypersetup{draft}
\begin{document}

\begin{frontmatter}
\maketitle

\begin{abstract}
Breakthrough Listen is a 10-yr initiative to search for signatures of technologies created by extraterrestrial civilizations at radio and optical wavelengths. Here, we detail the digital data recording system deployed for Breakthrough Listen observations at the 64-m aperture CSIRO Parkes Telescope in New South Wales, Australia. The recording system currently implements two modes: a dual-polarization, 1.125~GHz bandwidth mode for single beam observations, and a 26-input, 308~MHz bandwidth mode for the 21-cm multibeam receiver. The system is also designed to support a 3~GHz single-beam mode for the forthcoming Parkes ultra-wideband feed.  In this paper, we present details of the system architecture, provide an overview of hardware and software, and present initial performance results.

\end{abstract}

 \end{frontmatter}

\section{Introduction}

We are entering a golden age of astrobiology. 
In the last two years, Earth-like planets have been discovered around some of the nearest stars to Earth: Proxima Centauri \citep{Escude:2016}, Ross 128 \citep{Bonfils:2017}, and Luyten's star \citep{AstudilloDefru:2017}, and are now known to be common around sun-like stars \citep{Petigura:2013} as well as far more numerous small stars \citep{Dressing:2015}. Upcoming missions---the Transiting Exoplanet Survey Satellite (TESS\footnote{\href{https://tess.gsfc.nasa.gov}{tess.gsfc.nasa.gov}}), the James Webb Space Telescope (JWST\footnote{\href{https://www.jwst.nasa.gov}{jwst.nasa.gov}}), and the European Extremely Large Telescope (E-ELT\footnote{\href{https://www.eso.org/sci/facilities/eelt/}{www.eso.org/sci/facilities/eelt/}}), among numerous others---will further inform our understanding of exoplanet formation, abundance, composition, and atmospheres.
.

While exoplanets are now known to be common, the prevalence of life beyond Earth remains undetermined. The ongoing Search for Extraterrestrial Intelligence (SETI) seeks to place constraints on the presence of technologically-capable life in the Universe via detection of artificial `technosignatures' \citep{Cocconi1959, Drake1961}. Early radio SETI searches were limited by instrumentation to narrow bandwidths \citep{Drake1961}; as a result, searches were often concentrated around so-called `magic frequencies' \citep[e.g.][]{Blair:1992,Gindilis1993}, such as near the 21-cm neutral hydrogen emission line. As the capabilities of telescopes and signal processing systems increased exponentially, in lock-step with Moore's law, the instantaneous bandwidth over which radio searches could be conducted increased from kilohertz \citep[e.g.][]{Cohen:1980, Werthimer1985} to gigahertz \citep[e.g.][]{Macmahon:2017}. The search for coherent emission has also expanded into infrared \citep[e.g.][]{WrightS2001,WrightS:2015} and optical \citep[e.g.][]{Reines2002, Howard2004, Tellis:2015, Veritas:2016} wavelengths.  \cite{Tarter2001} provides an excellent review of SETI searches up to the turn of the century.

This combination---rapid progress in exoplanet science and exponential increase in digital capabilities---has motivated a new search for technologically-capable civilizations beyond Earth: the Breakthrough Listen initiative \citep[BL,][]{Isaacson:2017, Worden:2017}. Launched in July 2015, BL is a 10-year scientific SETI program to systematically search for artificial electromagnetic signals from beyond Earth, across the electromagnetic spectrum. In its first phase, BL is using the Robert C. Byrd Green Bank 100-m radio telescope in West Virginia, USA; the CSIRO Parkes 64-m radio telescope in New South Wales, Australia; and the 2.4m Automated Planet Finder at Lick Observatory in California, USA. A core component of BL is the installation of next-generation signal processing systems at the Parkes and Green Bank telescopes, to allow for wide bandwidth ($\sim$GHz) spectroscopy with ultra-fine frequency and time resolution ($\sim$Hz, $\sim$ns), and the ability to capture voltage-level data products to disk for phase-coherent searches.

In this article, we report on the Breakthrough Listen data recorder system for Parkes (\dr); the BL digital systems for the Green Bank telescope are detailed in \citet{Macmahon:2017}. The \dr\ system is designed to digitize, record and process the entire $\sim$4.3\,GHz aggregate bandwidth provided by the the 21-cm multibeam, and similarly the entire $\sim$1~GHz of bandwidth provided by the single beam conversion system. The \dr\ will also support a $0.7 - 4.0$\,GHz receiver, scheduled to be installed in Q2 2018.  Both these and subsequent bandwidth figures are for dual polarizations.

\subsection{CSIRO Parkes 64-m Telescope}
\label{sec:csiro}

The CSIRO Parkes radio telescope is a 64-m aperture, single dish instrument located north of the town of Parkes, New South Wales, Australia (32$^\circ$59'59.8''S, 148$^\circ$15'44.3''E). Over the period October 2016--2021, a quarter of the annual observation time of the Parkes 64-m radio telescope has been dedicated to the BL program. The Parkes telescope will be used to survey nearby stars and galaxies, along with a survey of the Galactic plane at 21-cm wavelength \citep{Isaacson:2017}.

The Parkes telescope has a suite of dual-polarization single-beam receivers that may be installed at its prime focus; details of receivers in use with the digital system explicated in this paper are presented in \reftab{tab:receivers}. Parkes is also equipped with a multibeam receiver operating over 1.23--1.53\,GHz, which consists of an array of 13 cryogenically cooled dual-polarization feeds \citep{Staveley1996}.

The single beam receivers at Parkes share a common signal conditioning and downconversion system  that provides up to 1\,GHz of bandwidth to digital signal processing `backends'. This conversion system (Parkes Conversion System, PCS, in \reffig{fig:architecture}) is user-configurable, and distributes the downconverted signals from the receiver to the digital backends installed at the telescope. As of writing, the three main backends are: the HI-Pulsar system (HIPSR), used for 21-cm multibeam observations \citep{Price2016}; the DFB4 digital filterbank  system, used for single beam observations in both spectral and time domain modes; and a data recorder system for use with Very Long Baseline Interferometry (VLBI). As the BL science program required $\sim$Hz resolution data products over the full bandwidth of the receivers, these existing backends were not appropriate (see \reftab{tab:survey-speed}). 

Several SETI experiments have been conducted previously with Parkes. \citet{Blair:1992} conducted a 512-channel, 100\,Hz resolution survey of 176 targets (nearby stars and globular clusters) at 4.46\,GHz. As part of the Project Phoenix initiative, \citet{Tarter1996} observed 202 solar-type stars, covering 1.2--3\,GHz. These observations were carried out with a digital spectrometer system with 1\,Hz resolution, 10\,MHz instantaneous bandwidth. This spectrometer was also used by   \cite{Shostak:1996}, who observed three 14~arcminute fields within the Small Magellanic Cloud, covering 1.2--1.75\,GHz.  \citet{Stootman:2000} implemented a `piggyback' SETI spectrometer, SERENDIP South, on two beams of the multibeam receiver; this 0.6\,Hz resolution, 17.6\,MHz bandwidth spectrometer was designed for commensal observations, however, no scientific outputs were published.

\begin{table}
	\caption{Parkes single beam receivers used in BL observations. }
	\label{tab:receivers}
	\centering
	\begin{tabular}{l c c c}
	\hline
	Receiver        & Frequency    & IF bandwidth & Sensitivity \\
	                & (GHz)        & (MHz)      & (Jy/K)     \\
	\hline
	\hline
	H-OH            & 1.2--1.8       & 500       & 1.2\\
	10cm            & 2.6--3.6       & 900       & 1.1\\
	Mars            & 8.1--8.5       & 500       & 1.7\\
	13mm            & 16.0--26.0     & 900       & 2.2\\
	21-cm MB        & 1.23--1.53     & 300       & 1.1\\
	\hline
	\end{tabular}
\end{table}

\subsection{Paper overview}

This paper provides an overview of the \dr~system for the Parkes 64-m telescope and is organized as follows. In \refsec{sec:architecture}, the overall system architecture is introduced. \refsec{sec:hardware} provides details of selected hardware; this is followed by sections detailing firmware (\refsec{sec:firmware}) and software (\refsec{sec:software}). In \refsec{sec:results}, initial on-sky results are presented, along with system verification and diagnostics. The paper concludes with discussion of scientific capabilities and future plans.

\section{System Overview}\label{sec:architecture}

\begin{figure*}
\begin{center}
\includegraphics[width=1.8\columnwidth]{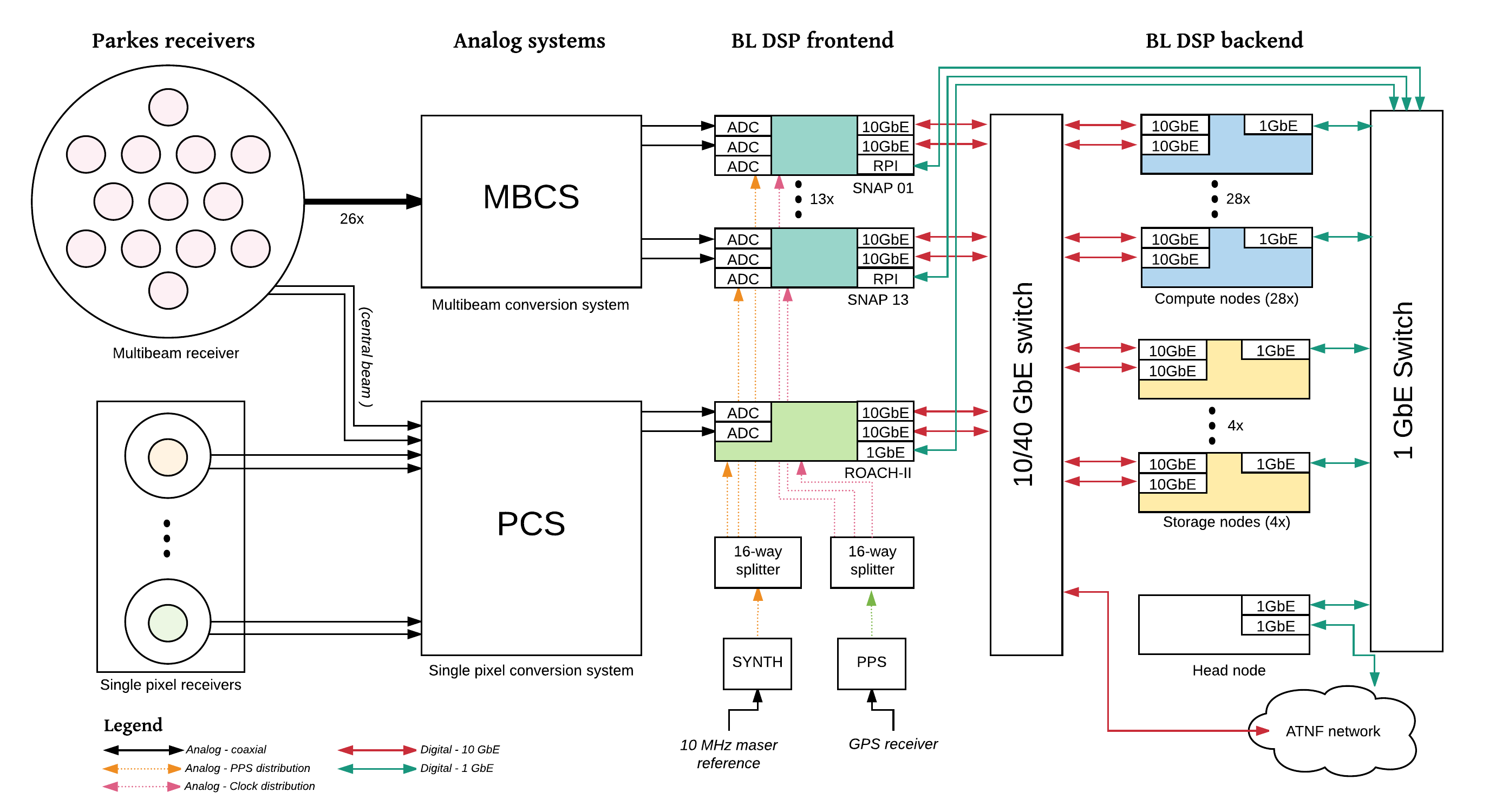}
\protect\caption{Block diagram of the Breakthrough Listen data recorder system architecture at Parkes. The BL DSP backend is shared between the single-beam and mutibeam DSP frontend systems. \label{fig:architecture}
}
\end{center}
\end{figure*}

\dr~is a heterogeneous digital signal processing (DSP) system that digitizes the signal from the telescope, records data to disk, and performs DSP tasks for data analysis and reduction. The overall architecture of the \dr~ system is comparable with that of the BL data recorder system installed at Green Bank \citep{Macmahon:2017}: a field-programmable gate array (FPGA) signal processing `frontend' is connected via high-speed Ethernet to a signal processing `backend' consisting of compute servers equipped with graphics processing units (GPUs). System monitor and control is carried out via a `head node' server, which also interfaces with the Parkes telescope control systems to collect observation metadata. A set of 4 storage-only servers are also installed; a block diagram of the \dr~is shown in \reffig{fig:architecture}.

Unlike the Green Bank installation, where the VEGAS FPGA system \citep{Prestage2015}	 was repurposed, new hardware was installed for the DSP frontend (\refsec{sec:hardware}). As Parkes has both a single beam and multibeam conversion system, two distinct FPGA frontends have been commissioned. The first utilizes a single CASPER ROACH-II FPGA board\footnote{\url{https://casper.berkeley.edu/wiki/ROACH-2_Revision_2}} (\refsec{sec:roach2}) with a 5~Gsample/s ADC (ADC5G\footnote{\url{https://casper.berkeley.edu/wiki/ADC1x5000-8}}), while the second utilizes a set of CASPER SNAP FPGA boards\footnote{\url{https://casper.berkeley.edu/wiki/SNAP}} with in-built ADCs (\refsec{sec:snap}).

During observations, \dr~records critically-sampled voltage data to disk at up to 750~MB/s per compute node. All data reduction is performed immediately after observations, with reduced data products and a subset of raw voltage data archived onto storage servers.
The two FPGA frontends are connected via 10~GbE to a shared compute backend. During observations, the FPGA frontend applies coarse channelization to the input signals and distributes channels between the backend compute servers via packetized Ethernet. A data capture code (\refsec{sec:software}) receives the Ethernet packets, arranges the packets into ascending time order, and writes their data payloads into an in-memory ring buffer. The ring buffer contents are then written to disk, along with telescope metadata. 

\subsection{Deployment timeline}

The \dr~system was deployed in several phases. In September 2016, a system comprised of two compute nodes and the ROACH-II was deployed. This allowed recording of up to 375 MHz bandwidth, and was used for initial commissioning tests. A further four compute nodes and one storage server were installed December 2016, which expanded recording capability to 1.125~GHz bandwidth; this bandwidth allowed recording of the full instantaneous bandwidth provided by the single-beam conversion system. In June 2017, the SNAP boards were installed, and the system was expanded to its full complement of 27 compute nodes plus 4 storage nodes. Observations with the multibeam system commenced in October 2017.

\subsection{HIPSR commensal mode}

During \dr~commissioning, the Parkes telescope control systems were updated to support commensal observations with the HIPSR \citep{Price2016} and \dr~systems. This upgrade was motivated by the desire to perform real-time searches for fast radio bursts (FRBs) and pulsars during BL observations with the multibeam receiver. Since commissioning, HIPSR's capabilities have been extended to support new features such as capturing polarization information and improving real-time FRB detection capabilities \citep{Petroff:2015, Keane:2018}. Indeed, HIPSR is itself an upgrade of the Berkeley Parkes Swinburne Recorder \citep[BPSR,][]{McMahon:2008, Keith:2010}. Nevertheless, HIPSR does not record voltage-level data products due to hardware limitations. Commensal observations therefore allow for voltage-level data products to be captured around an FRB, enabling new science.

The HIPSR real-time FRB detection system emails candidate events to a list of collaborators for verification. In the case of a bona-fide FRB, the BL observer in charge is contacted to ensure that voltage data around the event is retained, and relevant follow-up observation calibration routines can be conducted as appropriate. This strategy has already been successfully used to capture the voltage data from FRB 180301 \citep{Price:2018atel}; see \refsec{sec:frb}.

\section{Digital Hardware}\label{sec:hardware}

The \dr ~consists of off-the-shelf Supermicro servers, Ethernet networking hardware, and FPGA processing boards developed by the Collaboration for Astronomy Signal Processing and Electronics Research \citep[CASPER,][]{Hickish2016}. This hardware is detailed in the following subsections.

\subsection{ROACH-II FPGA board}\label{sec:roach2}

The CASPER Reconfigurable Open-Architecture Compute Hardware version 2 (ROACH-II) is an digital signal processing platform centered around a Xilinx Virtex-6 series FPGA (SX475T). The ROACH-II was custom designed for radio astronomy applications, and is deployed in a variety of use cases; see \cite{Hickish2016} for an overview.  In \dr , each ROACH-II is equipped with two ADC5G daughter cards, and a 10~GbE SFP+ Ethernet module. This configuration is the same as the VEGAS system installed at Green Bank, as used by \dr 's sister instrument. As the bandwidth provided by the Parkes downconversion system is 1~GHz, a single ROACH-II board is sufficient to digitize the entire available bandwidth.

The ROACH-II board is equipped with a PowerPC microprocessor, with which monitor and control of the board is conducted. The PowerPC runs a lightweight variant of the Debian Linux operating system, provided as part of the ROACH-II board support package. The PowerPC allows for the FPGA to be reprogrammed remotely; after programming, control registers on the FPGA are presented to the PowerPC via a memory-mapped interface. The ROACH-II is controlled over the network via the use of the KATCP\footnote{Karoo Array Telescope Control Protocol, \href{https://github.com/ska-sa/katcp}{github.com/ska-sa/katcp}} protocol.

For signal digitization, the \dr\ ROACH-II is equipped with dual CASPER ADC5G daughter cards, as described in \cite{Patel2014}. The ADC5G is an 8-bit, 5~Gsample/s ADC module for the ROACH-II, and is based upon the e2V EV8AQ160 chip\footnote{\href{http://www.e2v.com/products/semiconductors/adc/ev8aq160/}{www.e2v.com/products/semiconductors/adc/ev8aq160/}}. The ADC5G runs at up to 5~Gsample/s, providing up to 2.5\,GHz of digitized bandwidth. In \dr, the ADCs are configured to digitize a 1.5\,GHz band. The ADC5G part was selected as it has suitable dynamic range, covers the full bandwidth of the Parkes downconversion system, and furthermore is thoroughly characterized and in widespread use within radio astronomy.

\subsection{SNAP board}\label{sec:snap}

The Smart Network ADC Processor, or SNAP \citep{Hickish2016}, is an FPGA platform primarily designed for the Hydrogen Epoch of Reionization array \citep[HERA, ][]{Deboer2017}. The SNAP board is centered around a Xilinx Kintex-7 FPGA and three on-board Hitite HMCAD1511 ADC chips\footnote{\href{http://www.analog.com/en/products/analog-to-digital-converters/ad-converters/hmcad1511.html}{www.analog.com/en/products/analog-to-digital-converters/ad-converters/hmcad1511.html}}. The HMCAD1511 is a multi-input, 8-bit ADC chip that digitizes 4 streams at up to 250~Msamples/s. Alternatively, the ADC cores may be interleaved to digitize 2 inputs at 500~Msample/s, or 1 input at 1000~Msample/s. In the \dr , the 1000~Msample/s mode is used exclusively. 

\begin{figure}
\includegraphics[width=0.95\columnwidth]{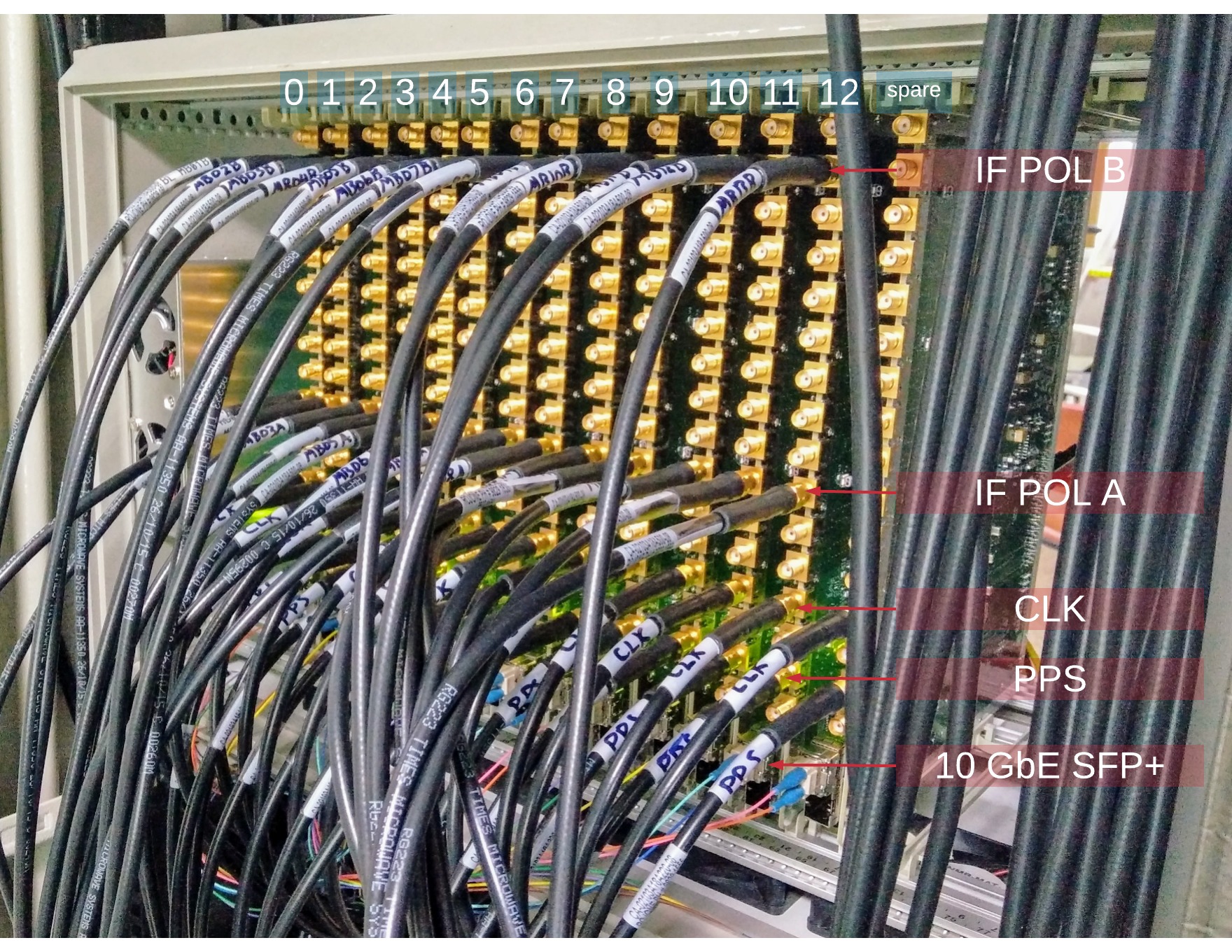}
\caption{\label{fig:snap-front} Front view of SNAP boards as installed at Parkes in VME-style chassis, showing analog connections. A total of 14 boards (one per beam plus spare) are housed within the chassis.
}
\end{figure}

\begin{figure}
\includegraphics[width=0.95\columnwidth]{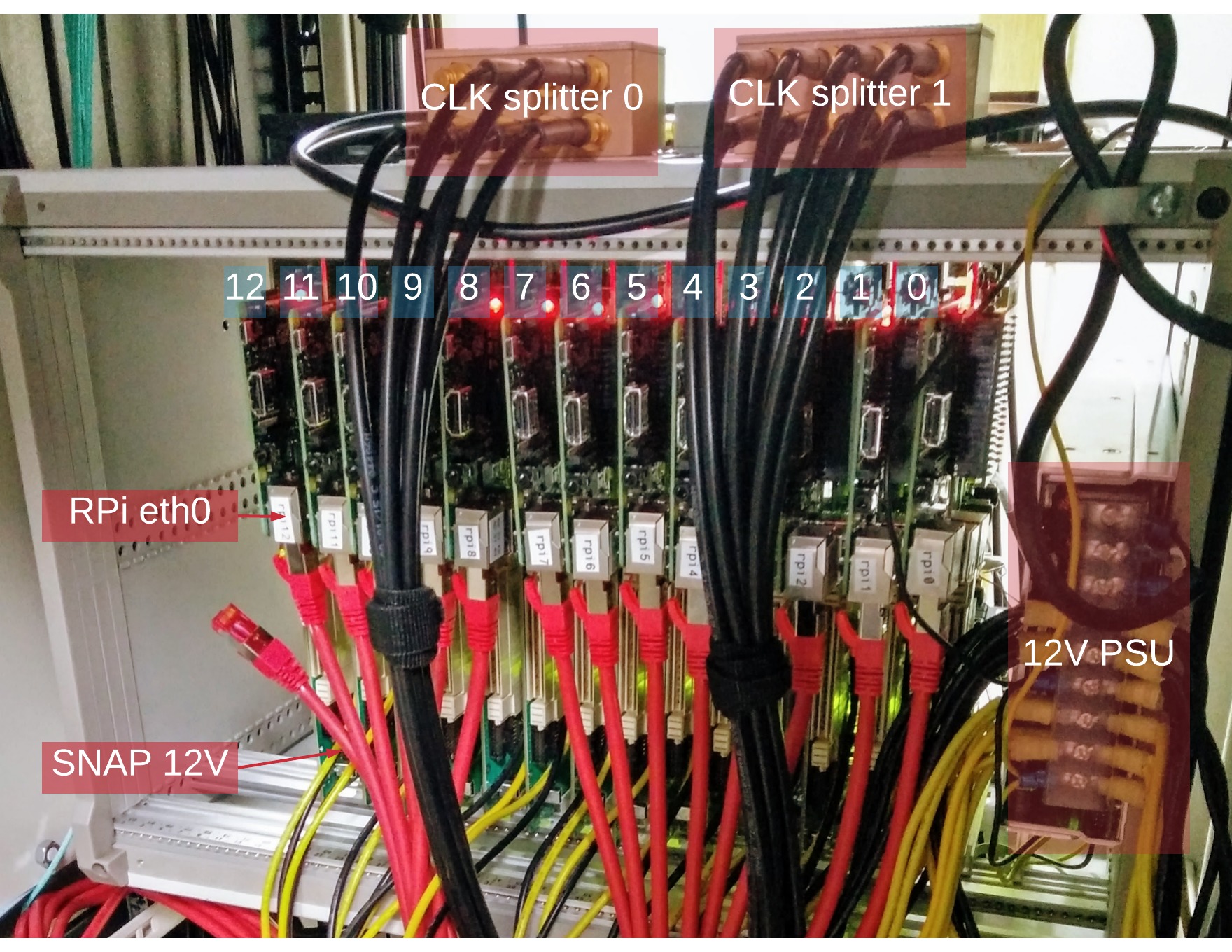}
\caption{\label{fig:snap-back} Rear view of SNAP boards as installed at Parkes. A Raspberry Pi 2 (RPi) is connected to each SNAP board via the GPIO pins; each RPi has a 1~GbE connection for remote monitor and control of its corresponding SNAP board.}
\end{figure}

In \dr , the 14 constituent SNAP boards are mounted vertically in a 8U VME-style chassis (\reffig{fig:snap-front}). Also housed within this chassis is a 12\,V power supply (Meanwell SP-480-12), which is shared between the boards. Monitor and control of each SNAP board is handled by a Raspberry Pi 2 Model B\footnote{\href{http://www.raspberrypi.org}{www.raspberrypi.org}} (RPi) processor board connected to  via a 40-pin GPIO header (\reffig{fig:snap-back}). The GPIO header provides the RPi board with 5\,V power. Each RPi runs a KATCP server, via which the FPGA may be programmed and software registers may be configured. 

In the \dr ~installation, each SNAP board accepts a reference clock, pulse-per-second, and two signal inputs (\reffig{fig:snap-front}). Each board digitizes a dual-polarization pair of signals from the 13 beams within the multibeam receiver; the third ADC is unused. The 14th SNAP within the VME chassis is configured as a spare in case of board failure.

%$%% UP TO HERE %%%$%$%$%$%%%%

\subsection{Clock synthesizer and pulse-per-second}

Each ADC requires an external reference frequency standard `clock signal' to be provided. Additionally,  the FPGA clock is derived from the ADC clock signal on both the SNAP and ROACH-II platforms. Clock signals for the ROACH-II ADC5G and SNAP HMCAD1511 ADC are generated by a two-channel Valon 5008 frequency synthesizer, at 3000\,MHz and 896\,MHz, respectively. For enhanced stability, the  synthesizer is locked to a 10\,MHz reference tone provided by an on-site maser source. To avoid frequency drift between SNAP boards, a single clock signal  is distributed via a power divider network. 

A nanosecond pulse-per-second (PPS) signal derived from the Global Positioning System (GPS) is also distributed to each board. The PPS allows for boards to trigger data capture on the rising edge of the PPS, synchronizing the boards to within one clock cycle ($\sim 5$\,ns). 

\subsection{Compute servers}

\begin{figure}
\includegraphics[width=0.95\columnwidth]{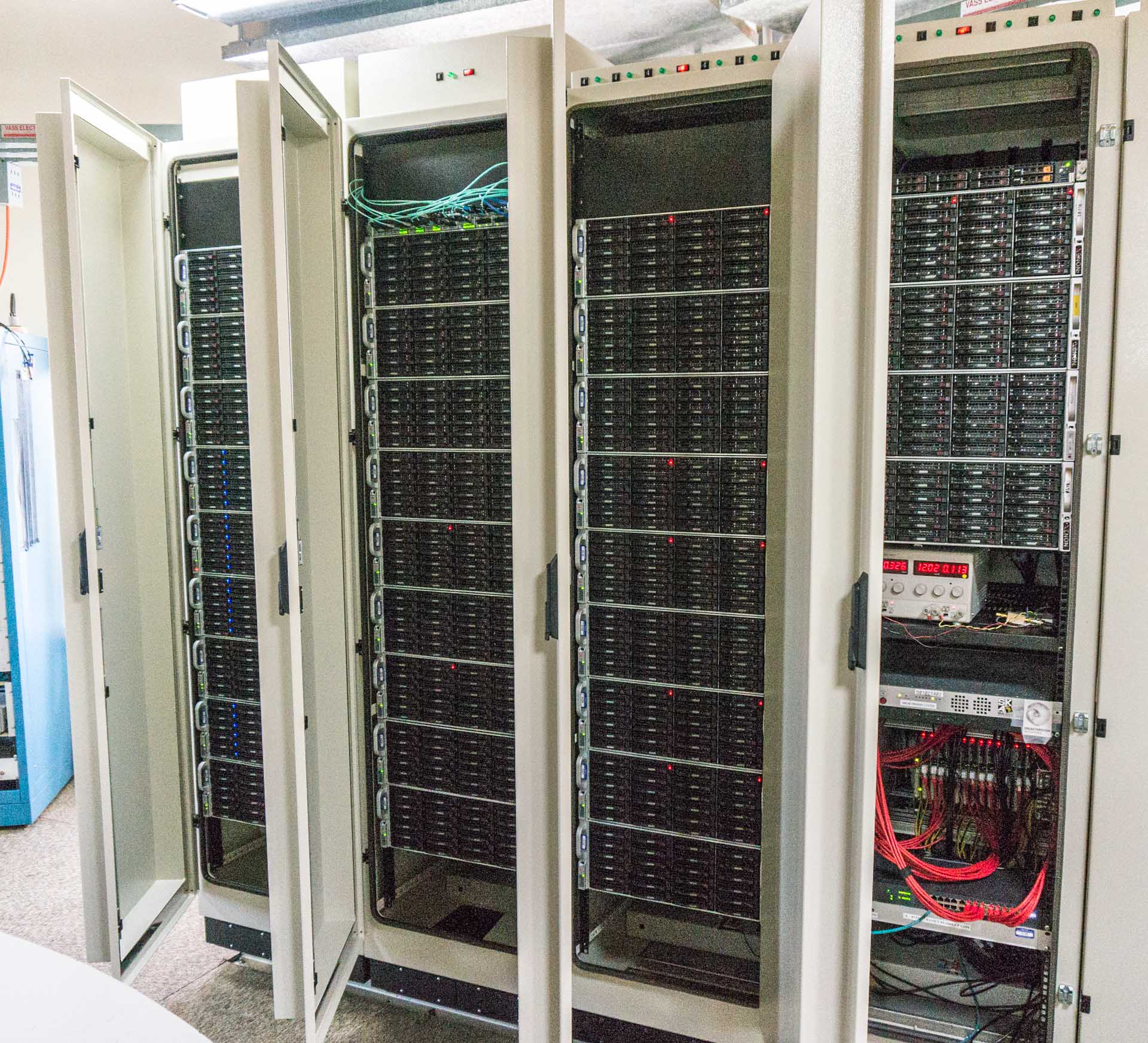}
\caption{\label{fig:rack} Rack front pic.}
\end{figure}

\begin{table}
	\caption{\dr ~compute node configuration.}
	\label{tab:compute}
	\centering
	\begin{tabular}{l r}
	\hline
	Chassis         & Supermicro 4U server \\
	Motherboard     & Supermicro X10DRi-T4+ \\
	CPU             & 2x Intel Xeon E5-2620v4 \\
	GPU             & NVIDIA GTX 1080 \\
	Memory          & 64GB DDR4 RAM   \\
	Hard drives     & 24x Toshiba 5TB 7200 rpm \\
	RAID controller & LSI SAS 3108 MegaRAID \\
	10 Gb Ethernet  & Mellanox Connect-X3 2xSFP+ \\
	\hline
	\end{tabular}
\end{table}

The \dr ~compute servers have two purposes: during observations, to capture voltage data from the FPGA frontend to disk, and between observations, to reduce voltage data into time-averaged dynamic spectra for future data analysis and for archiving. Voltage-level data products for signals of interest may also be stored.

For our purposes, this approach is advantageous over real-time data reduction as it allows use of data reduction algorithms that run slower than real time and act on file objects. Nonetheless, a real-time data analysis pipeline to run in parallel with raw data capture is under active development. 

For intercompatibility with our Green Bank systems, we selected 4U Supermicro systems (\reftab{tab:compute}), built to specification by the Australian distributor Xenon Systems.

A total of 27 compute nodes (26 + 1 spare) are installed in the \dr ~system (\reffig{fig:rack}). The overall bandwidth that can be recorded to disk is proportional to the number of compute servers; each server may capture up to 187.5~MHz of bandwidth to disk at 8 bits per sample for two inputs (750~MB/s disk write). For the full 26-node system, an aggregate 4.875~GHz of dual-polarization bandwidth may be captured. Each node is equipped with 24\,x~5\,TB hard drives, configured into two RAID 5 partitions (single parity) with 2 hot spares.

\subsection{Storage servers}

\begin{table}
	\caption{\dr ~storage node configuration.}
	\label{tab:storage}
	\centering
	\begin{tabular}{l r}
	\hline
	Chassis         & Supermicro 4U server \\
	Motherboard     & Supermicro X10DRi-T4+ \\
	CPU             & 2x Intel Xeon E5-2620v4 \\
	GPU             & None (on-board only)\\
	Memory          & 64\,GB DDR4 RAM   \\
	Hard drives     & 36x Toshiba 6\,TB 7200 rpm\\
	RAID controller & LSI SAS 3108 MegaRAID \\
	40 Gb Ethernet  & Mellanox Connect-X3 1xQSFP+ \\
	\hline
	\end{tabular}
\end{table}

The four storage servers are similar in configuration to the compute servers (\reftab{tab:storage}), but with a chassis that houses 36 disks, and 40~Gb Ethernet adaptors.  A discrete GPU is not included. 

A full complement of 36\,x~6\,TB hard drives are installed per server (216~TB total). Each storage server is configured with three RAID 5 arrays of 11 disks formatted with XFS; the remaining 3 disks are configured as global hot spares. This configuration provides 165~TB of storage per node. 

\subsection{Ethernet interconnect}

The SNAP boards, ROACH-II, compute and storage nodes are connected together via a 10\,Gb Ethernet network. Nodes are connected via a Arista 7050QX-32 switch that has thirty-two 40\,GbE QSFP+ ports; QSFP+ ports are broken out into four 10\,GbE SFP+ ports using Fiberstore 850\,nm fiber optic breakout cables (FS48510) with a QSFP+ transceiver on one end (FS17538) and four SFP+ transceivers (FS33015) on the other ends. 

The \dr\ head node is connected to the observatory-managed network via a 1\,GbE connection; a secondary 1\,Gb Ethernet interface on the head node connects the head node to the \dr\ private internal network, the interconnect for which is provided by a Netgear S3300 1\,Gb Ethernet switch. Shielded category 6A cables are used throughout the \dr\ network, so as to conform to observatory requirements aimed at minimizing self-generated radio interference.

\subsection{Power and cooling}

The \dr ~system is installed on the second floor of the telescope tower, directly underneath the steerable dish structure. The \dr ~hardware is installed within four RF-tight cabinets with standard air-cooled 19-inch racks. Each rack may draw a maximum of 10~kW of 240V AC power. The total power budget of the \dr ~is 19.6~kW, a breakdown of the power budget is given in \reftab{tab:power}. 

Power usage is at maximum during data reduction on the GPU cards; the cards are rated to draw up to 180~W when in operation. In practice, the GPUs draw $\sim$40~W when not in use, and   <120~W when running our reduction codes.

Cooled air is delivered at the bottom of the racks and is pulled through to the top by fans. Signals enter the rack via either BNC-type coaxial feedthroughs or radio-tight fiber optic feedthroughs.

\begin{table}
	\caption{\dr ~power budget, not including cooling.}
	\label{tab:power}
	\centering
	\begin{tabular}{l c c c}
	\hline
	Item           & Quantity  & Power/unit & Total \\
	               &     -     & (W)        & (kW)  \\
	\hline
	\hline
	ROACH-II       & 1         & 100       & 0.1   \\
	SNAP boards    & 14        & 25        & 0.35   \\
	Arista 7050QX  & 1         & 1200      & 1.2   \\
    Networking: other & 1      & 500       & 0.5   \\
    Clock + PPS    & 1         & 200       & 0.2   \\
	Head node      & 1         & 400       & 0.4   \\
	Compute nodes  & 27        & 550       & 14.85  \\
	Storage nodes  & 4         & 500       & 2.0   \\
	\hline
	               &           &           & \textbf{19.6} 
	\end{tabular}
\end{table}

\section{FPGA Firmware}\label{sec:firmware}

\subsection{Single beam firmware (ROACH-II)}

For single-beam observations, the \dr\ uses the same 512-channel VEGAS Pulsar Mode firmware as used in the Green Bank system; we refer the reader to \citet{Macmahon:2017} for details. Briefly, the firmware digitizes two inputs at 3.0~Gsample/s, then applies a 512-channel polyphase filterbank to produce `coarse' channels. Subsets of 64 channels (187.5\,MHz) are then sent over 10\,Gb Ethernet to the compute nodes for further processing. In normal operation, the boards are configured to output six of the eight 187.5\,MHz sub-bands (1.125\,GHz total), to better match the $\sim$1\,GHz of usable bandwidth from the telescope's downconversion system.

\subsection{Multibeam firmware (SNAP)}

For observations with the multibeam receiver, a set of 13 SNAP boards are used to digitize, channelize, and output selected channels over 10\,Gb Ethernet. The SNAP boards run a shared firmware, designed and compiled with the JASPER toolflow \citep{Hickish2016}, using Xilinx Vivado 2016.4 and Mathworks MATLAB/Simulink  2016b. 

Each board accepts a polarization pair from the 26 IFs provided by the multibeam downconversion system. Programmable registers on the firmware allow for each board to be uniquely identified and configured; each SNAP outputs data over 10\,GbE to a different pair of compute nodes. The board firmware:

\begin{itemize}
\item Nyquist-samples the input signals at 8 bits with a sample rate of 896\,MHz. 
\item Coarsely channelizes the data using a 256-point, 16-tap polyphase filterbank (PFB) running on the FPGA. The resultant 128 channels have a full-width half maximum (FWHM) of 3.5\,MHz, with a complex-valued fixed-point bitwidth of 18 bits. 
\item Requantizes the PFB output down to 8 bits. A runtime-configurable equalization gain is used allow tuning for optimal quantization efficiency.
\item  Buffers up selected channels (runtime configurable).
\item Converts selected channels into UDP Ethernet packets and outputs them over 10\,GbE.
\end{itemize}
The packetized output data is sent via the SNAP board's 10\,GbE SFP+ connector to the compute nodes for further processing. 

In normal operation, the boards are configured to output 88 channels (channels 20--107), 308\,MHz resultant bandwidth. Each compute node receives 44 contiguous channels (154~MHz) for a single SNAP board.

\subsubsection{Diagnostic shared memory}

To ensure the power levels are appropriate, sampled data may be captured at several points in the firmware design: directly after digitization, at the output of the PFB, and after requantization to 8 bits. This capture is done via the use of block RAMs on the FPGA that are accessible to the Raspberry Pi via memory mapping.

\subsubsection{Packet format and data rate}
The data payload of each UDP packet consists of 4 contiguous channels, with 512 frames (i.e. time samples) per packet. From slowest to fastest varying, data are arranged as (frame, channel, polarization, real/imaginary sample). The data payload is preceded by a 64-bit header, consisting of a 48-bit packet identifier, 8-bit channel identifier, and 8-bit beam identifier (ranging from 0--12). The packet identifier increases monotonically with time, to allow reconstruction of the voltage stream during depacketization.

The total packet size is 8200\,B; as such, jumbo frames support is required on all 10~GbE  interfaces (the maximum transmission unit, MTU, must be increased to 8200\,B or greater). A minimum of 4 channels must be selected for output, corresponding to an output data rate of 448.4\,Mb/s per board. When configured for regular operations, 88 channels are output (308\,MHz bandwidth), and the corresponding data rate is 9.866\,Gb/s per board (128.253\,Gb/s aggregate).

\section{Software}\label{sec:software}

\subsection{Telescope integration}
The Parkes telescope is controlled by the observer through a system called \textsc{tcs} (Telescope Control System). \textsc{tcs} provides a user interface through which observing schedules may be run and telescope configurations applied.  The \dr ~system receives telescope metadata (pointing information, observing frequency setup, observer name, etc) from \textsc{tcs} via an Ethernet TCP\footnote{Transmission Control Protocol} socket. Metadata are communicated over this socket with simple plaintext keyword-value pairs, to which a plaintext response of \lq OK' is waited for before sending the next keyword. To enable commensal observations, the \textsc{tcs} system was upgraded to  broadcast metadata to both HIPSR and \dr~in parallel.

A server daemon running on the \dr ~headnode responds to the messages from \textsc{tcs} and stores these metadata to a \textsc{redis}\footnote{\href{https://redis.io}{redis.io}} database (\reffig{fig:metadata}). Where required, server-specific metadata are computed from information provided by \textsc{tcs}, such as channel frequency values. This daemon also triggers the start and stop of data capture when relevant commands are received from \textsc{tcs}.

\subsection{HASHPIPE data capture}

The \textsc{hashpipe}\footnote{\href{https://github.com/david-macmahon/hashpipe}{github.com/david-macmahon/hashpipe}} software package is used to capture UDP packets from the DSP frontend and write these data to disk. \textsc{hashpipe} is written in C and implements an in-memory ring buffer into which data are shared between processes, as detailed in \citet{Macmahon:2017}.

\begin{figure}
\includegraphics[width=1\columnwidth]{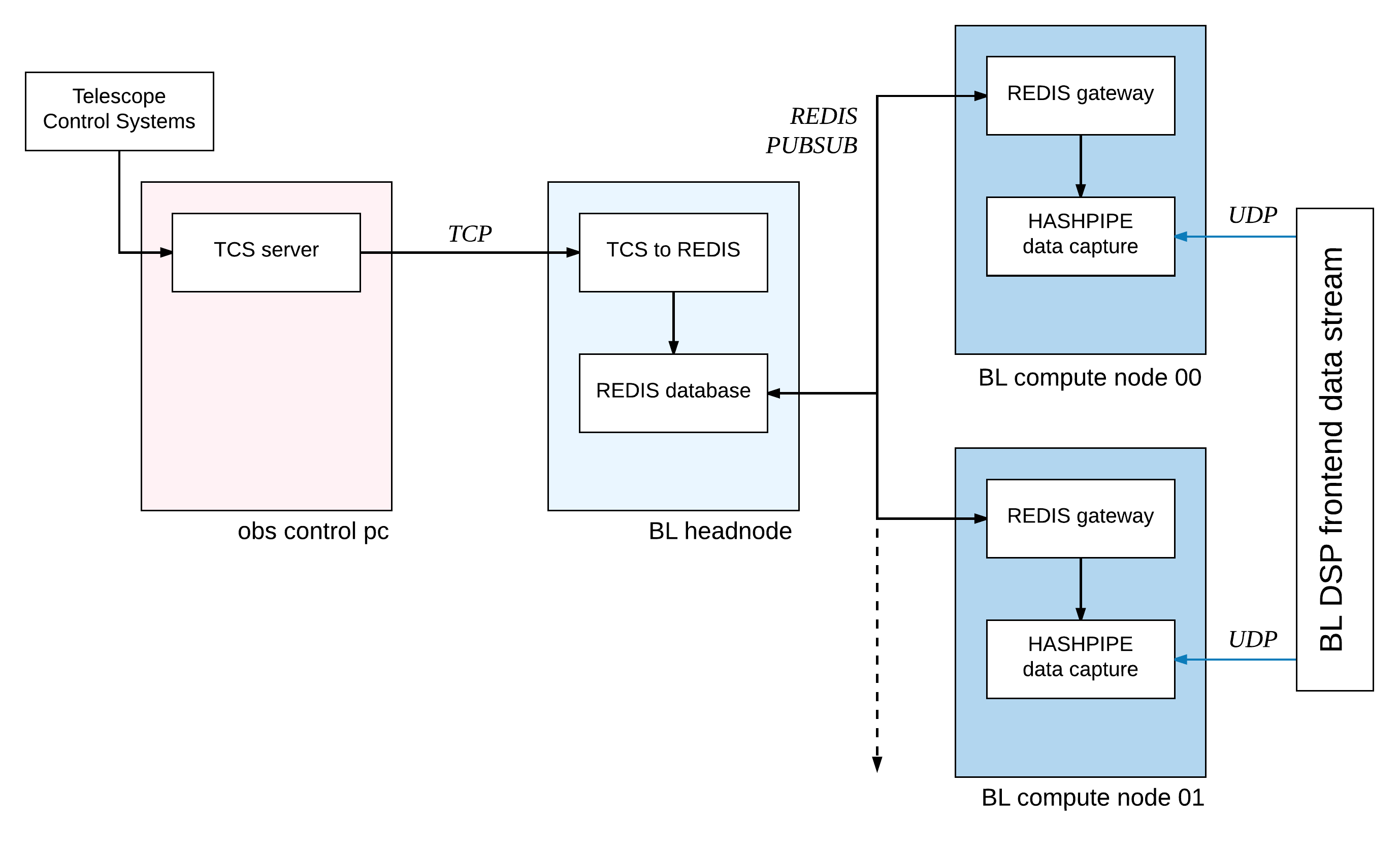}
\protect\caption{\label{fig:metadata} Block diagram showing metadata propagation from Parkes telescope control systems to \dr ~data capture processes.
}
\end{figure}

In both single-beam and multibeam mode, a copy of the \textsc{hashpipe} pipeline is launched on each compute server. Packetized data from the FPGA boards are sent via 10\,GbE to the compute nodes (\reffig{fig:metadata}); metadata are collected from the \textsc{redis} database running on the headnode. The \textsc{hashpipe} pipeline writes these metadata and data to files in the GUPPI raw format \citep{Ford:2010}. 

Different invocations of the \textsc{hashpipe} pipeline are required for the two modes of operation. For the single-beam mode, only 6 of the compute nodes are required to capture 1.125\,GHz of digitized bandwidth from the FPGA frontend; each node captures a subset of 64 channels (187.5\,MHz each). For multibeam mode, 26 compute nodes are used, each capturing a subset of forty-four 3.5\,MHz channels (154\,MHz per node), over 1.228--1.536\,GHz.

\subsection{GPU data reduction}

In order to produce power spectral density data products, we have implemented a GPU-accelerated data reduction pipeline called \textsc{bunyip}\footnote{\href{https://github.com/ucberkeleyseti/bunyip}{github.com/ucberkeleyseti/bunyip}} using the  \textsc{bifrost} framework \citep{Cranmer:2017}. This code, in order of operation:

\begin{itemize}
	\item Reads data from file in GUPPI raw format. 
	\item Performs a fast Fourier Transform (FFT). This operation is performed on the GPU device, using the NVIDIA \textsc{cufft} library.
	\item Squares the output, and integrates the data in GPU device memory.
	\item Combines polarizations to form Stokes I data.
	\item Writes output data to file in either Sigproc filterbank\footnote{\href{http://sigproc.sourceforget.net}{sigproc.sourceforget.net}} format or HDF5\footnote{\href{http://www.hdfgroup.org}{www.hdfgroup.org}} format. 
\end{itemize}

For multibeam observations, the pipeline produces three products with different time and frequency resolution pairs: (3.37\,Hz, 19.17\,s); (3.42\,kHz, 0.60\,s); (437.5\,kHz, 292.57\,$\mu$s). For single beam observations, resolutions are: (2.79\,Hz, 18.25\,s); (2.86\,kHz, 1.07\,s); (366\,kHz, 349.53\,$\mu$s). Data formats will be detailed further in Lebfosky et. al. (in prep).

\section{Verification and Results}\label{sec:results}

\subsection{Single-beam observations}

The single-beam system uses the same digitizer and FPGA firmware as the Green Bank install; as such, we followed the same verification procedures detailed in \citet{Macmahon:2017}. We first verified data throughput through Ethernet UDP capture tests, and benchmarking disk write speed. We then confirmed the probability distribution for digitized on-sky data were Gaussian, as expected for noise-dominated astrophysical signals. By injection of test tones, we confirmed our frequency metadata was correct to better than the width of our highest-resolution data product (2.79\,Hz). 

Here, we present a sampling of results that are illustrative of system performance.

\subsubsection{Proxima centauri}

\begin{figure}
\includegraphics[width=1\columnwidth]{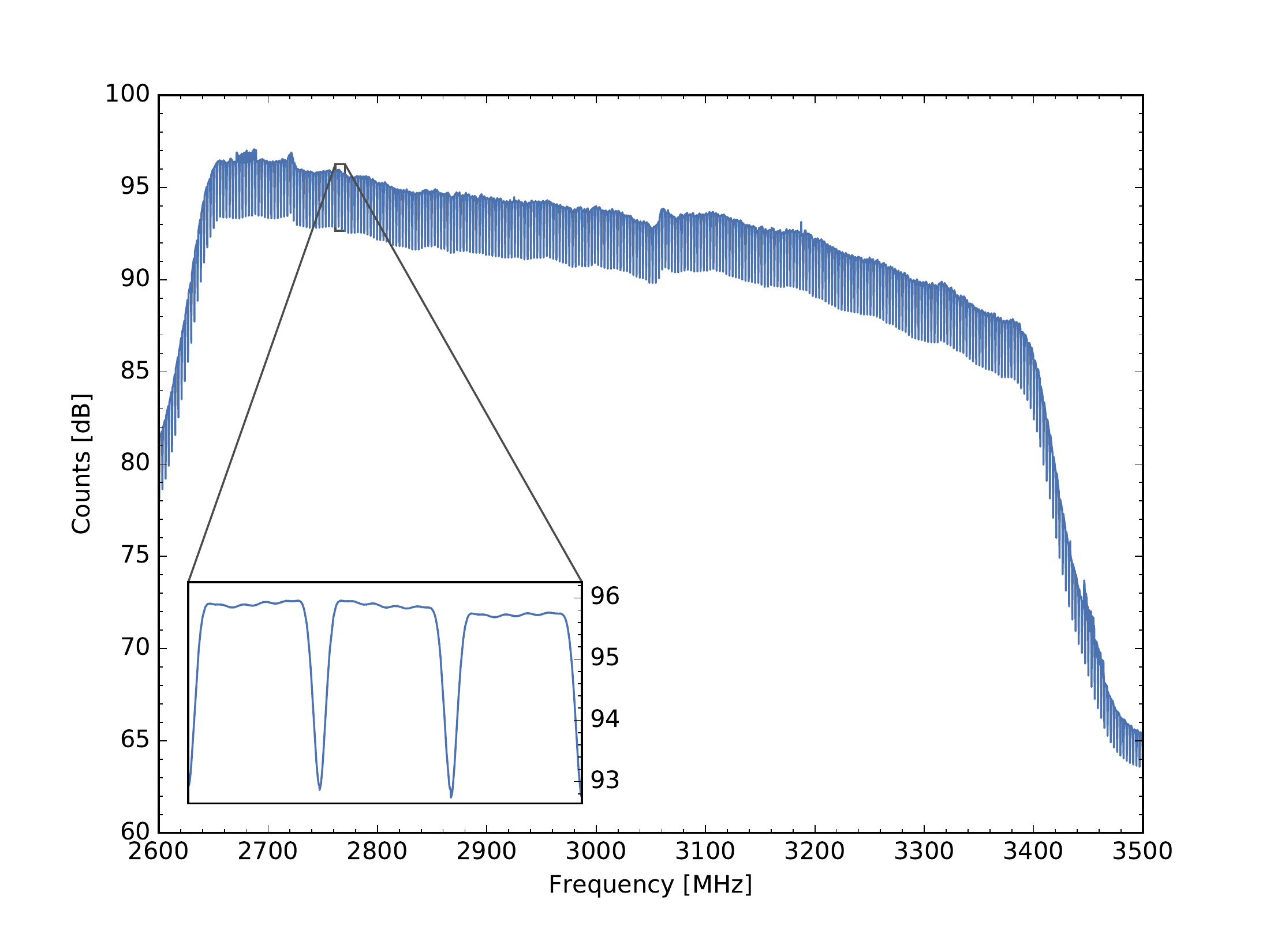}
\protect\caption{\label{fig:proxima-bp} Uncalibrated mid-resolution (327680 channels over 937.5~MHz) spectrum on Proxima Centauri, using the 10-cm receiver on 2017 January 20. The inset shows a zoom over three coarse channels, showing the characteristic filter shape.}
\end{figure}

At a distance of 4.2 light years, Proxima Centauri is the nearest star to Earth, after the Sun. In August 2016, the discovery of an Earth-sized exoplanet orbiting Proxima Centauri within the so-called `habitable zone' was announced by \citet{Escude:2016}. This makes Proxima Centauri an interesting candidate for SETI observations.

We observed Proxima Centauri for five minutes on 2017 January 20, using the 10-cm receiver. Five compute nodes were used to record 187.5\,MHz of voltage data each, spanning between 2.6--3.5375\,GHz (937.5\,MHz total). Stokes-I spectra (2.86\,kHz, 1.07\,s) from the observation were created using our GPU-accelerated pipeline. 

The 5-minute time-averaged spectrum from Proxima Centauri is shown in \reffig{fig:proxima-bp}. The inset shows the characteristic filter shape of the coarse channels; the filter shape of the telescope's downconversion system can be seen at the band edges. These data are typical of 10-cm observations. Our observations and data analysis of Proxima Centauri and other nearby star targets as listed in \citet{Isaacson:2017} are ongoing, and will be the subject of a future publication.

\subsubsection{J0835$-$4510 (Vela) single pulses}

\begin{figure}
\includegraphics[width=1.1\columnwidth]{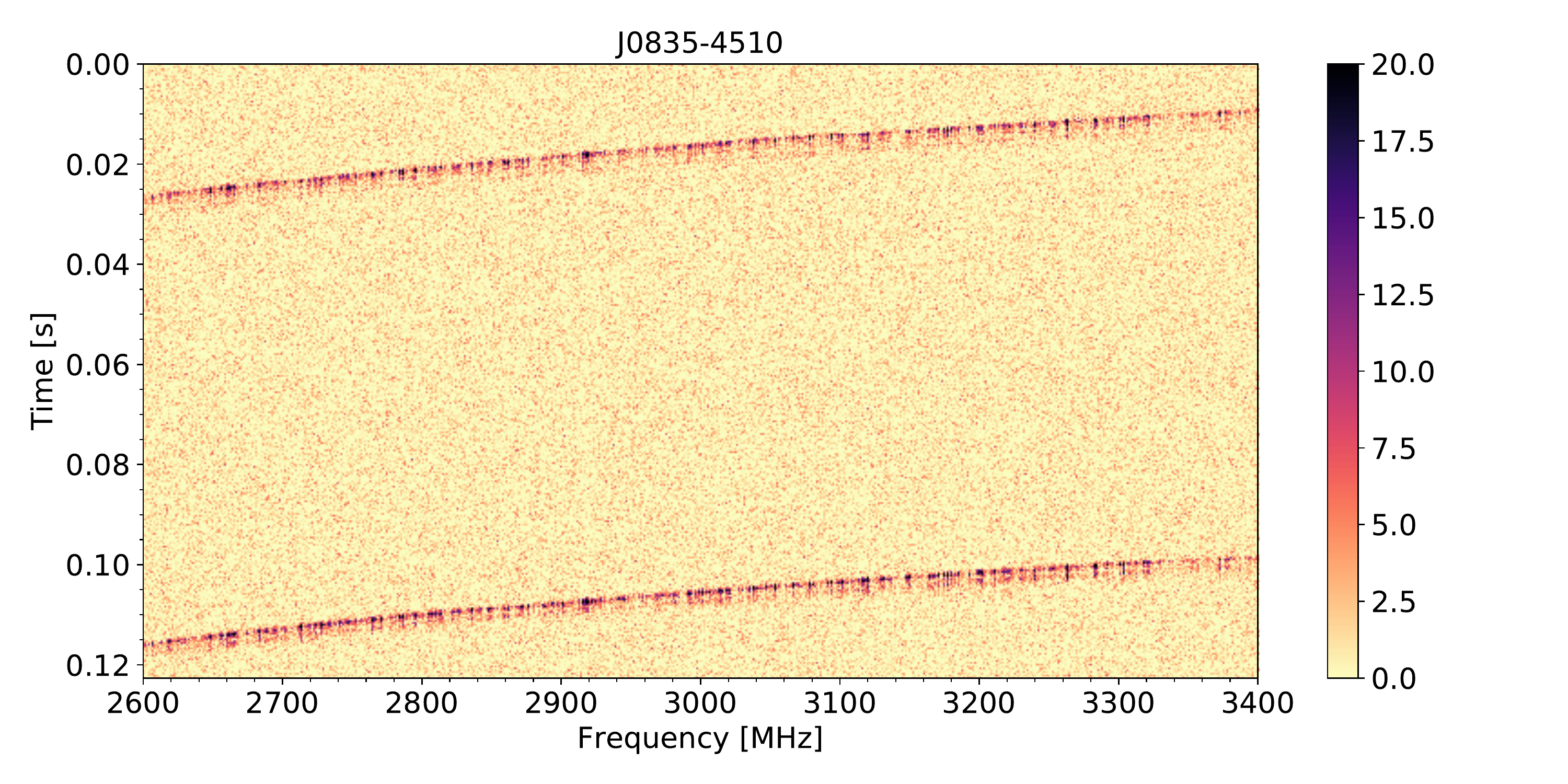}
\protect\caption{\label{fig:vela} Two broadband pulses at 10-cm wavelength from the Vela pulsar (PSR J0835-4510, MJD 57941.295), after bandpass removal. Colour scale is flux in Jy, calibrated using an (approximate) 38.5~Jy system equivalent flux density for the 10-cm receiver. 
}
\end{figure}

PSR\,J0835$-$4510 --- the Vela Pulsar --- is the brightest pulsar in the Southern sky. PSR J0835$-$4510 has a period of 89.33\,ms, mean flux density of 1100\,mJy at 1.4\,GHz, and dispersion measure of 67.97\,pc\,cm$^{-3}$ \citep{Manchester:2005}.

We observed PSR\,J0835$-$4510 for 5 minutes using the 10-cm receiver and \dr~system, on UTC 2017 July 7 at 07:04. The data were recorded across five compute nodes, reduced into Stokes-I filterbank files with 366\,kHz, 349.53\,$\mu$s resolution, then combined to form a single filterbank file spanning 2.6--3.4\,GHz. 
Two broadband pulses from these observations are shown in \reffig{fig:vela}. The resultant period, DM, and other parameters for PSR\,J0835$-$4510 are consistent with previously known values.

\subsubsection{Voyager 2}

Voyager 2 is a NASA space probe launched on 1977 August 20. Currently, at a distance of over 116~AU from Earth, Voyager is one of the most distant human-made objects, but its narrowband telemetry signal can still be detected. This makes observations of Voyager 2 an excellent test of the \dr~capabilities. 

\begin{figure}
\includegraphics[width=1\columnwidth]{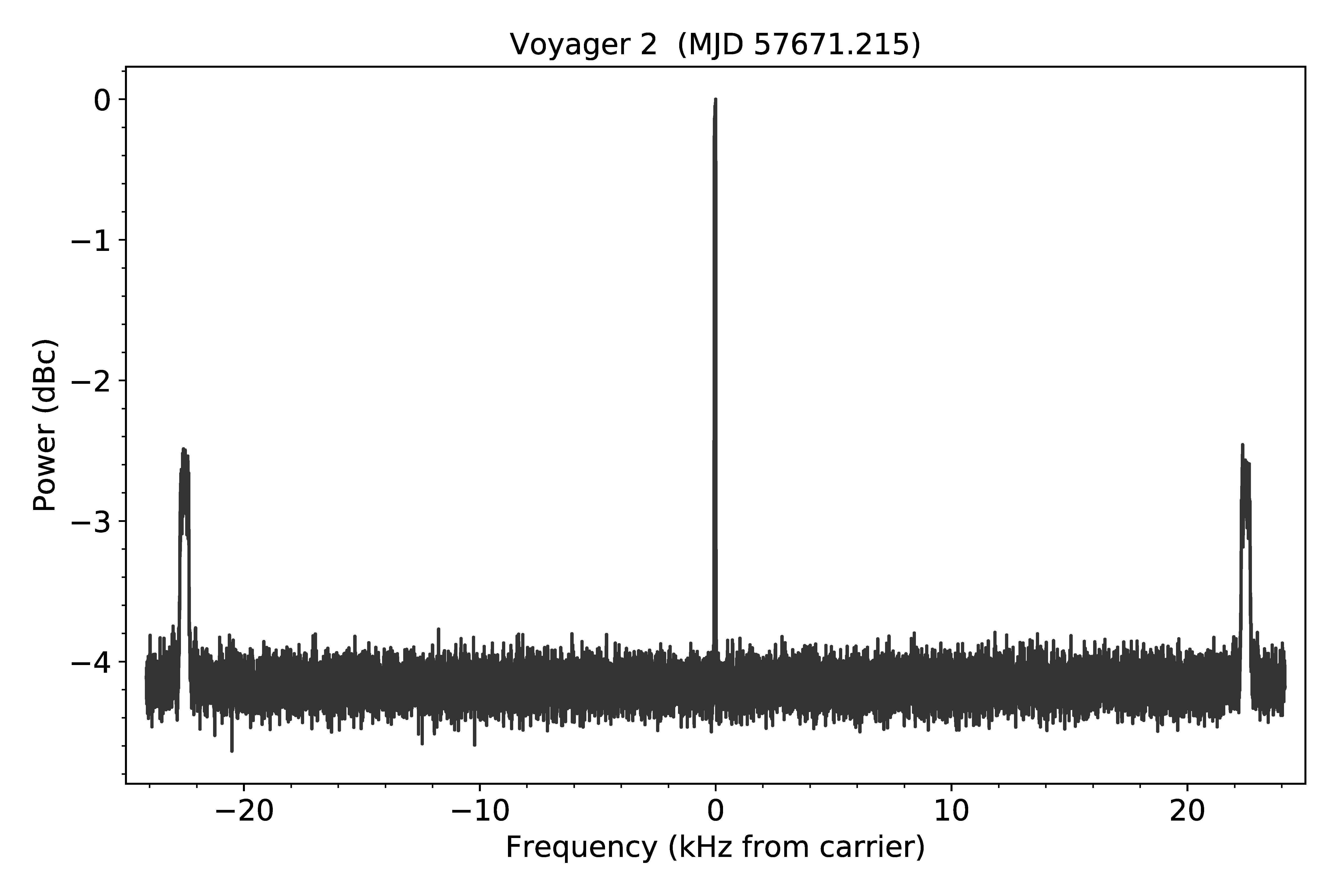}
\protect\caption{\label{fig:voyager} Voyager 2 space mission telemetry signal at UTC 09:37 2016 October 10, detected using the Parkes Mars receiver and the \dr. 
}
\end{figure}

We observed Voyager 2 on UTC 2016 October 10 at 09:37, using the Parkes Mars receiver (8.1--8.5\,GHz). The J2000 ephemeris at observation, retrieved from the NASA HORIZONS website\footnote{\href{https://ssd.jpl.nasa.gov/horizons.cgi}{ssd.jpl.nasa.gov/horizons.cgi}}, was (19:58:36, $-$57:18:53.5). Using \dr, we recorded 5 minutes of data, and created Stokes I filterbank files with 2.79\,Hz resolution. \reffig{fig:voyager} shows the detected telemetry signal; the carrier, upper and lower sideband are clearly visible. Here, we have not corrected for doppler broadening of the signal.

\subsection{Multibeam observations}

\begin{figure*}
\includegraphics[width=2\columnwidth]{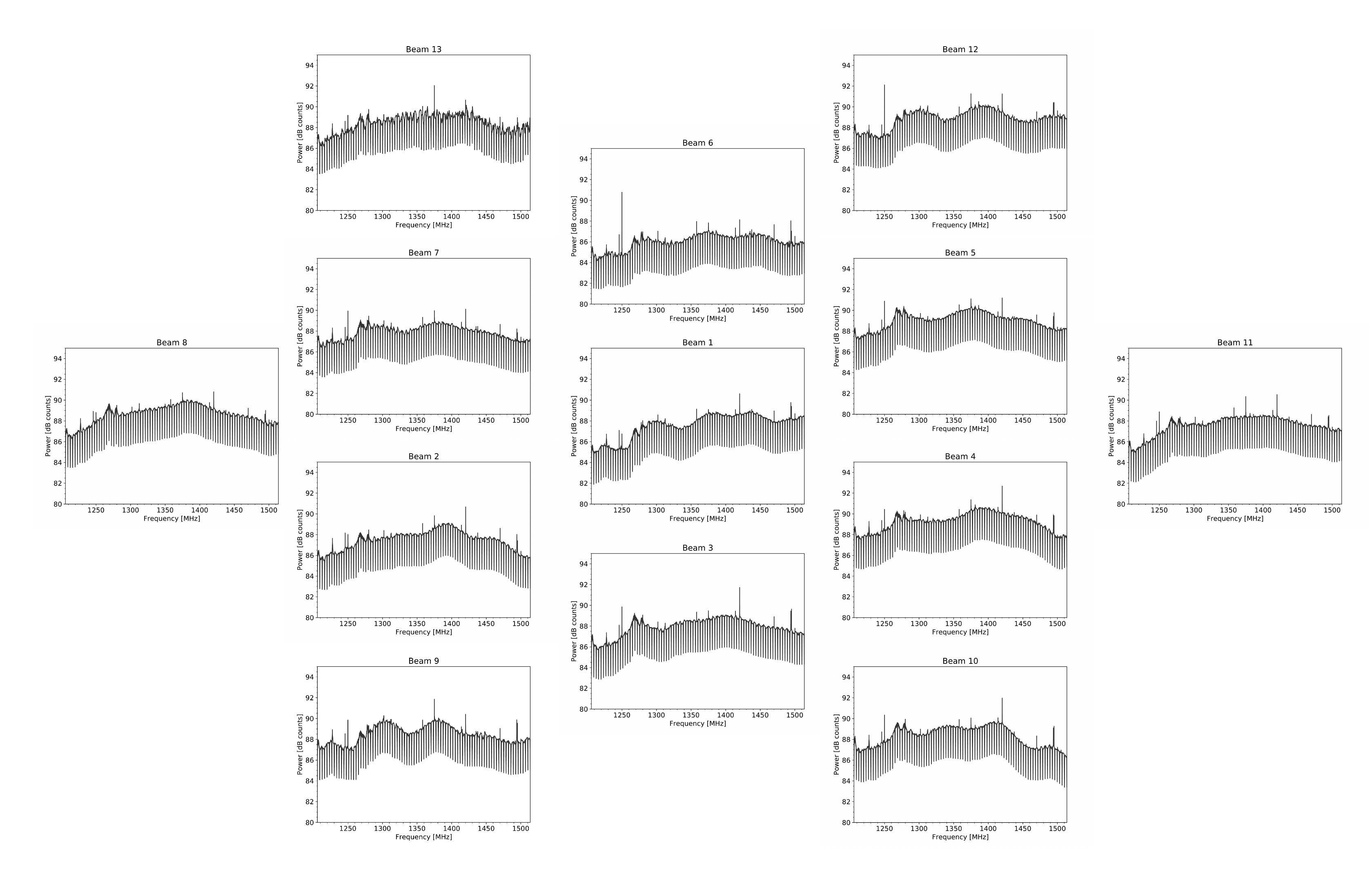}
\protect\caption{\label{fig:multibeam} Example Stokes-I spectra from \dr~showing bandpass for the 13 beams of the 21-cm multibeam receiver; plots are laid out in the same hexagonal manner as the receiver. Here, beam 01 is centred on globular cluster NGC~1851 (J2000 coordinates RA 5:14:06, DEC $-$40:02:47).
}
\end{figure*}

The \dr~system receives a total of 26 inputs from the multibeam conversion system. A set of configurable attenuators in the conversion system allows for power levels at the ADCs to be matched; we aim for input powers of -20~dBm, and run the ADCs with a 8x digital gain such that input RMS levels are between 8--16 counts. This level provides high quantization efficiency (better than 99\%), while leaving >5 bits of headroom for radio interference. 

\reffig{fig:multibeam} shows example spectra for all 13 beams, taken from a 5-minute observation of globular cluster NGC~1851 (RA 5:14:06, DEC $-$40:02:47), on UTC 2018 February 27 at 05:35; globular clusters were identified as potential `cradles of life' by \cite{DiStefano:2016}. As in Fig.~\ref{fig:proxima-bp}, the coarse bandpass filter shape can be seen. Sources of radio interference (RFI) can be identified as common to all beams.

\subsubsection{FRB 180301}\label{sec:frb}

\begin{figure}
    \includegraphics[width=1.0\linewidth]{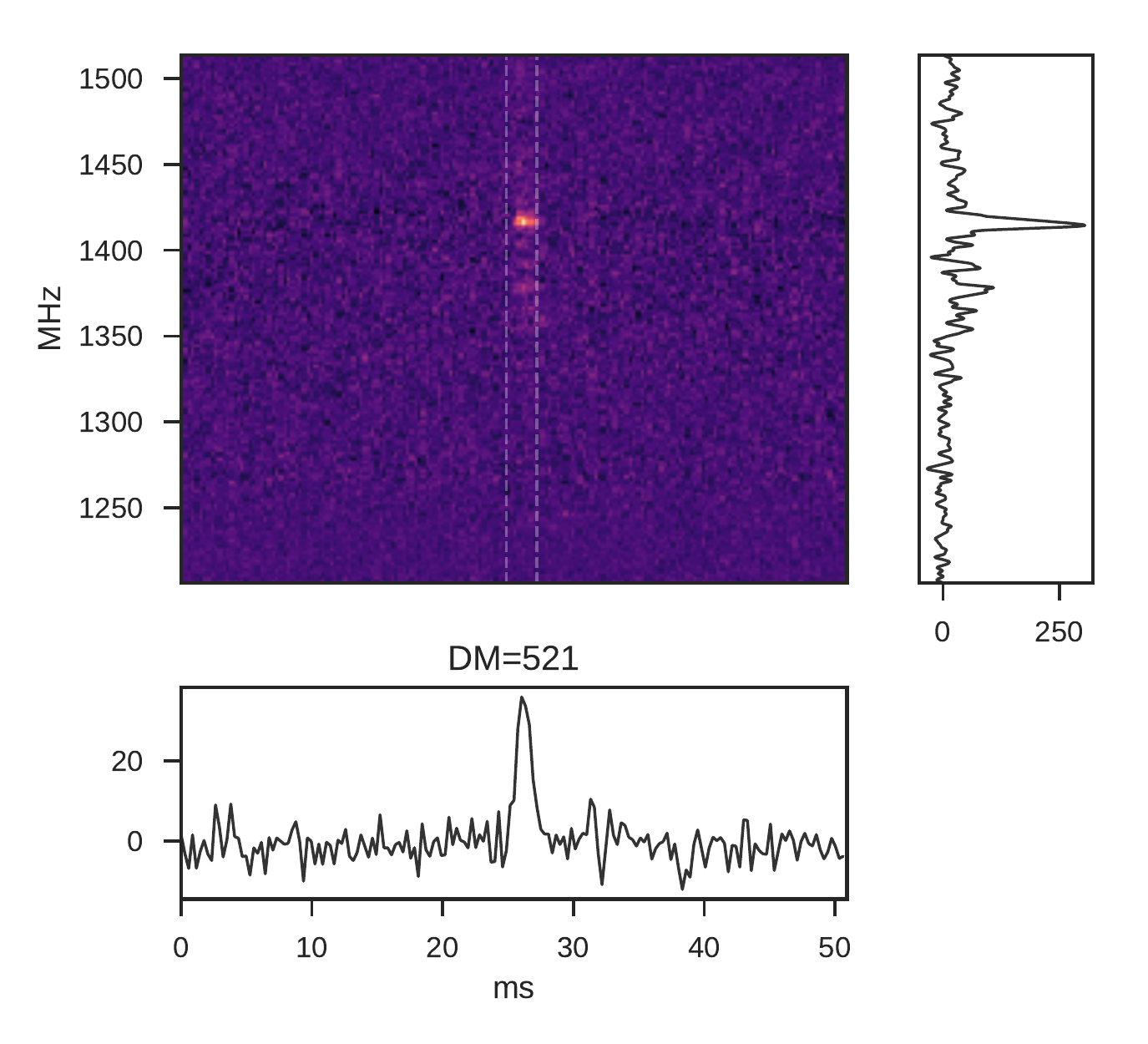}
    \caption{Incoherently dedispersed spectrum of FRB 180301.  Dedispersed time
    series using a dispersion measure of 521~pc~cm$^{-3}$ plotted at the bottom.
    The average frequency structure of the pulse, within the dashed lines of the
    dynamic spectrum, is plotted on the right. Frequency structure and time
    series in arbitrary flux units.
    }
    \label{fig:FRB180301}
\end{figure}

On 2018 March 01 at UTC 07:34, a fast radio burst (FRB) was detected by the
commensal HIPSR real-time system during BL observations with the multibeam
receiver \citep{Price:2018atel}. An email trigger
was broadcast by the HIPSR system to the SUPERB collaboration \citep{Keane:2018}
and BL for verification, and the observer in charge was notified within minutes so that calibration routines could be performed. The burst, FRB\,180301 (Figure \ref{fig:FRB180301}), was detected approximately six degrees below the Galactic
plane, at J2000 RA 06:12:40, with an uncertainty of $\pm20$s, and declination +04:33:40, within an uncertainty of $\pm20''$.  The burst was detected in beam 03 of the receiver, with
an observed signal-to-noise ratio of 16, a  $\sim$0.5 Jy peak flux density,
$\sim$3~ms burst width and dispersion measure (DM) of $\sim$521 pc cm$^{-3}$.

The \dr~recorded the voltage data for this burst from each of the 13 beams of the receiver, all of which have been saved for analysis. These voltage-level products allow for coherent dedispersion of the FRB, measurement of its  rotation measure, polarization properties, and localization via inter-beam correlation. Analyses of FRB\,180301 are ongoing, and will be the subject of a future publication (Price et. al., in prep). 

\section{Discussion}

\subsection{SETI survey speed}

In \citet{Enriquez:2017}, various figures of merit are discussed for SETI surveys; for comparison of an instrument's intrinsic capability, the survey speed figure of merit (SSFM) is the most salient. The SSFM quantifies how fast a telescope can reach a target sensitivity, $S_{\rm{min}}$, for a given (narrowband) observation:
\begin{equation}
{\rm SSFM}\propto\frac{\Delta\nu_{{\rm obs}}}{{\rm SEFD}^{2}\delta\nu},
\end{equation}
where $\Delta\nu_{\rm{obs}}$ is the instantaneous bandwidth of the instrument,  $\delta\nu$ is the channel bandwidth, and SEFD is the System Equivalent Flux Density. 

The parameters $\delta\nu$ and $\Delta\nu_{\rm{obs}}$ are set by the capabilities of the digital systems (although constrained by the receiver), whereas the SEFD is determined by the telescope's collecting area and system temperature. As \dr~records voltages, the minimum channel resolution is constrained only by the observation length ($\delta\nu\propto t^{-1}$). \reftab{tab:survey-speed} gives a comparison of \dr~to other digital systems at Parkes; in comparison to the Project Phoenix \citep{Tarter1996} and SERENDIP South \citep{Stootman:2000} digital systems, the ratio $\Delta\nu_{\rm{obs}} / \delta\nu$ is over $10^4$ larger. For this comparison, the channel resolution at maximum instantaneous bandwidth is used. Note that for a FFT-based spectrometer, the ratio $\Delta\nu_{\rm{obs}} / \delta\nu$ is equal to the total number of channels.  

\begin{table}
	\caption{Comparison of channel bandwidth and instantaneous bandwidth---that is, aggregate $\Delta\nu_{\rm{obs}}$ over all beams---for digital backends at Parkes.}
	\label{tab:survey-speed}
	\centering
	\begin{tabular}{l c c c}
	\hline
	Digital        & $\Delta\nu_{\rm{obs}}$  & $\delta\nu$ & $\Delta\nu_{\rm{obs}} / \delta\nu$ \\
	backend        & (MHz)        & (Hz)        &     \\
	\hline
	\hline
	Proj. Phoenix   & 10        & 1.0               &  $1\times10^7$ \\
    SERENDIP South  & 17.6      & 0.6               &  $3\times10^7$ \\
	DFB4            & 1024      & 125$\times10^3$   &  8192   \\
	HIPSR           & 5200      & 48.8$\times10^3$  &  106496 \\
	\dr             & 4004      & 0.005$^\dagger$   &  $8\times10^{11}$\\
	\hline
    \multicolumn{4}{l}{$^{\dag}$ Min. resolution for 30-minute observation}
	\end{tabular}
\end{table}

\subsection{Conclusions}

\dr~is a new digital system for the CSIRO Parkes 64-m telescope, that can record up to an aggregate of 8.624~GHz of 8-bit data, at data rates of up to 128~Gb/s. This places it as the second highest maximum data recording rate of any radio astronomy instrument, after its sister installation in Green Bank \citep[see Tab.~5 of][]{Macmahon:2017}. 

The \dr~system is being used to undertake targeted observations of nearby stars and galaxies, and a Galactic plane survey; details of the survey strategy may be found in \citet{Isaacson:2017}. These observations, and related data analyses, are ongoing.

Through commensal observations with the HIPSR real-time FRB system, voltage data for transient events may be captured and analyzed. This strategy allowed the capture of voltage data from FRB\,180301, which will be the subject of a future publication. The ability to capture voltage data opens opportunities for other ancillary science, which may be supported on a shared-risk basis.

In the coming months, a 0.7--4\,GHz receiver will be installed at the Parkes telescope. This receiver will be digitized in the focus cabin, and data sent via 10\,GbE to a GPU-based spectrometer that is under development by the ATNF. A copy of these data will also be multicast to \dr~for SETI science; only modest modifications to the \dr~data capture codes will be required to support this new functionality.

\section{Acknowledgements}
Breakthrough Listen is  managed by the Breakthrough Initiatives, sponsored by the \href{http://breakthroughinitiatives.org}{Breakthrough Prize Foundation}. The Parkes radio telescope is part of the Australia Telescope National Facility which is funded by the Australian Government for operation as a National Facility managed by CSIRO. VG would like to acknowledge NSF grant 1407804 and the Marilyn and Watson Alberts SETI Chair funds. This work makes use of hardware developed by the Collaboration for Astronomy Signal Processing an Electronics Research (CASPER). D. Price thanks Matthew Bailes and Willem van Straten for site visits and discussions.

\bibliographystyle{pasa-mnras}
\bibliography{references}

\end{document}